# INFLUENCE OF INTERFACE TRAPS AND ELECTRON-HOLE PUDDLES ON QUANTUM CAPACITANCE AND CONDUCTIVITY IN GRAPHENE FIELD-EFFECT TRANSISTORS

[*]G. I. Zebrev, E.V. Melnik, A.A. Tselykovskiy

Department of Micro- and Nanoelectronics, National Research Nuclear University MEPHI, 115409,

Kashirskoe sh., 31, Moscow, Russia

**Abstract**

We study theoretically an influence of the near-interfacial insulator traps and electron-hole puddles on the small-signal capacitance and conductance characteristics of the gated graphene structures. Based on the self-consistent electrostatic consideration and taking into account the interface trap capacitance the explicit analytic expressions for charge carrier density and the quantum capacitance as functions of the gate voltage were obtained. This allows to extract the interface trap capacitance and density of interface states from the gate capacitance measurements. It has shown that self-consistent account of the interface trap capacitance enables to reconcile discrepancies in universal quantum capacitance vs the Fermi energy extracted for different samples. The electron-hole puddles and the interface traps impact on transfer I-V characteristics and conductivity has been investigated. It has been shown that variety of widths of resistivity peaks in various samples could be explained by different interface trap capacitance values.

## 1. Introduction

Capacitance measurements provide important information about density of states of mobile and localized states at the Fermi energy in the 2D and quasi-2D systems. The former is connected with the electronic compressibility and is often referred to as quantum capacitance [1]. The latter is associated with the interface traps which are capable to change their occupancy with gate bias changes and have energy levels distributed throughout the insulator bandgap [2].

The concept of "quantum capacitance" was introduced by Luryi [1] in order to develop an equivalent circuit model for devices that incorporate a highly conducting two-dimensional (2D) electron gas. The quantum capacitance can be considered as a direct generalization of the "inversion layer capacitance" in the silicon MOSFETs to the case of strictly one-subband filling. The inversion layer ("quantum") capacitance plays rather minor role in the silicon FETs since it is very low in subthreshold operation mode and extremely high in above threshold strong inversion regime. In the former case the quantum capacitance in MOSFETs is masked by the parasitic interface trap and depletion layer capacitances connected in parallel in the equivalent electric circuit, and in the latter case it is insignificant due to the series connection with the gate insulator having typically lesser capacitances for high carrier densities in inversion layers. In fact the inversion layer capacitance in MOSFETs is only important in the very narrow region of weak inversion where it is commensurable with the oxide and depletion layer capacitances. Graphene based FETs bring in absolutely new state of affairs. Quantum capacitance in graphene has an absolute minimum at the charge neutrality point, which in itself is not so small at room temperatures even for ideal graphene ($\leq 10\,f\text{F}/\mu\text{m}^2$ at 300 K) and slowly increases as linear function of the Fermi energy. In all range of the Fermi energy it may be commensurable with the capacitance of parasitic interface traps which unavoidable occur at the interface due to chemical and/or structural disorder. The depletion layer is absent in GFETs and the interface traps capacitance is the only in parallel connection with the quantum capacitance in the equivalent electric circuit. It is well known that high density of interface traps suppress electric field effect in gated structures generally and degrades field-effect mobility in particular [3]. Therefore the role of fast interface traps in operation of graphene gated structure as a FET needs to be understood [4]. This leads to importance of experimental discernment of interface traps and quantum capacitances since the interface trap density (and capacitance) can vary in a wide range of values depending on purity and quality of the interface or the substrate.

[*] gizebrev@mephi.ru

Interface traps act in similar way on transfer I-V characteristics of graphene FETs. Distorting the dependence of the gate voltage on the chemical potential the interface trap buildup leads to reduction of transconductance (or, field-effect mobility) even at permanent scattering rate or true mobility. Electron-hole puddles in graphene are another consequence of presence of the near-interfacial charged defects need not to be easily rechargeable. Electron-hole puddles modify quantum capacitance and conductivity near the charge neutrality point increasing its minimum value. The observed minimums of small-signal C-V characteristics are also strongly influenced by electron-hole puddles. The aim of this work is to develop a regular procedure for separation of the interface and quantum capacitances based on experimental capacitance data and to investigate the influence of the interface traps and the electron-hole puddles on capacitance and conductivity characteristics of graphene field-effect transistors.

The paper is organizes as follows. Sec.2 is devoted to derivation of the expression for quantum capacitance in ideal graphene without disorder and interaction. The interface traps are briefly discussed in Sec.3. General electrostatics and specific electrostatic parameters are the topics of Sec.3 and 4. Analytic expressions for graphene sheet charge density vs gate voltage and channel and gate capacitances taking into account interface trap density have derived in Sec.6 and 7. Analysis of literature experimental data and a separation procedure are presented in Sec.8 and 9. Impact of electron-hole puddles on quantum capacitance and mean conductivity of inhomogeneous graphene is considered in Sec.10 and 11.

## 2. Quantum Capacitance in Graphene

The density of states in clean graphene for dispersion law $\varepsilon = v_0\sqrt{p_x^2 + p_y^2}$ is given by

$$g_{2D}(\varepsilon) = \frac{2\varepsilon}{\pi \hbar^2 v_0^2}\mathrm{sgn}\,\varepsilon = \frac{2|\varepsilon|}{\pi \hbar^2 v_0^2}, \qquad (1)$$

where $\hbar$ is the Plank constant, $v_0$ ($\cong 10^8$ cm/s) is the characteristic (Fermi) velocity in graphene.

Using the equilibrium Fermi-Dirac function $f_{FD}(\varepsilon-\mu)$ the electron density per unit area $n_e$ at a given chemical potential $\mu$ for nonzero temperature $T$ reads

$$n_e(\mu) = \int_0^{+\infty} d\varepsilon\, g_{2D}(\varepsilon) f_{FD}(\varepsilon-\mu) =$$
$$= \frac{2(k_BT)^2}{\pi \hbar^2 v_0^2}\int_0^{+\infty} du\, \frac{u}{1+\exp\left(u-\frac{\mu}{k_BT}\right)} = -\frac{2}{\pi}\left(\frac{k_BT}{\hbar v_0}\right)^2 Li_2\left(-e^{\frac{\mu}{k_BT}}\right), \qquad (2)$$

where T is absolute temperature, $k_B$ is the Boltzmann constant, $Li_n(x)$ is the poly-logarithm function of n-th order [5]

$$Li_n(z) = \sum_{k=1}^{\infty} z^k/k^n. \qquad (3)$$

Using electron-hole symmetry $g(\varepsilon)=g(-\varepsilon)$ we have similar relationship for the hole density $n_h$

$$n_h(\mu) = \int_{-\infty}^0 d\varepsilon\, g_{2D}(\varepsilon)(1-f_{FD}(\varepsilon-\mu)) = -\frac{2}{\pi}\left(\frac{k_BT}{\hbar v_0}\right)^2 Li_2\left(-e^{-\frac{\mu}{k_BT}}\right) = n_e(-\mu). \qquad (4)$$

Full charge density per unit area or the charge imbalance reads as

$$n_S \equiv n_e - n_h = \int_0^{+\infty} d\varepsilon\, g(\varepsilon)(f(\varepsilon-\mu)-f(\varepsilon+\mu)) = \frac{2}{\pi}\left(\frac{k_BT}{\hbar v_0}\right)^2\left(Li_2\left(-e^{-\frac{\mu}{k_BT}}\right)-Li_2\left(-e^{\frac{\mu}{k_BT}}\right)\right). \qquad (5)$$

Conductivity of graphene charged sheet is determined by the total carrier density

$$N_S = n_e + n_h = -\frac{2}{\pi}\left(\frac{k_BT}{\hbar v_0}\right)^2\left(Li_2\left(-e^{-\frac{\mu}{k_BT}}\right)+Li_2\left(-e^{\frac{\mu}{k_BT}}\right)\right), \qquad (6)$$

notice that $N_S(\mu = 0) = (\pi/3)(k_B T/\hbar v_0)^2$.

The channel electron density per unit area for degenerate system ($\mu \gg k_B T$) reads

$$n_S \cong \int_0^\mu d\varepsilon\, g_{2D}(\varepsilon) = \frac{\mu^2}{\pi \hbar^2 v_0^2}. \tag{7}$$

Performing explicit differentiation of Eqs.(2,4) one reads

$$\frac{dn_e}{d\mu} = \frac{2}{\pi} \frac{k_B T}{\hbar^2 v_0^2} \ln\left(1 + \exp\left(\frac{\mu}{k_B T}\right)\right), \qquad \frac{dn_h}{d\mu} = -\frac{2}{\pi} \frac{k_B T}{\hbar^2 v_0^2} \ln\left(1 + \exp\left(-\frac{\mu}{k_B T}\right)\right). \tag{8}$$

Exact expression for quantum capacitance of the graphene charge sheet may be defined as

$$C_Q \equiv \int_{-\infty}^{+\infty} g(\varepsilon)\left(-\frac{\partial f_0}{\partial \varepsilon}\right) d\varepsilon = \frac{ed(n_e - n_h)}{d\mu} = \frac{2}{\pi}\left(\frac{e^2}{\hbar v_0}\right)\frac{k_B T}{\hbar v_0} \ln\left(2 + 2\cosh\left(\frac{\mu}{k_B T}\right)\right). \tag{9}$$

### 3. Near-interfacial rechargeable oxide traps

It is widely known (particularly, from silicon-based CMOS practice) that the charged oxide defects inevitably occur nearby the interface between the insulated layers and the device channel. Near-interfacial traps (defects) are located exactly at the interface or in the oxide typically within 1-3 nm from the interface. These defects can have generally different charge states and capable to be recharged by exchanging carriers (electrons and holes) with the device channels. Due to tunneling exchange possibility the near-interfacial traps sense the Fermi level position in graphene. These rechargeable traps tend to empty if their level $\varepsilon_t$ are above the Fermi level and capture electrons if their level are lower the Fermi level.

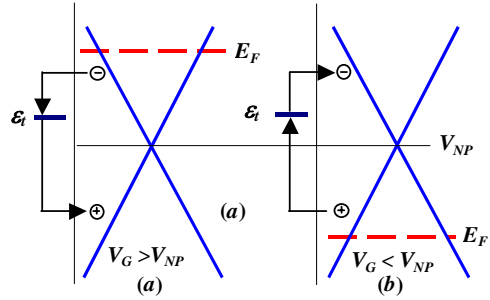

Fig. 1. Illustration of carrier exchange between graphene and oxide defects (a) filling; (b) emptying.

There are two types of traps – donors and acceptors. Acceptor-like traps are negatively charged in a filled state and neutral while empty ( - /0). Donor-like traps are positively charged in empty state and neutral in filled condition (0/+). In any case, the Fermi level goes down with an increase $V_G$ and the traps begin filled up, i.e. traps become more negatively charged (see Fig. 1). Each gate voltage corresponds to the respective position of the Fermi level at the interface with own "equilibrium" filling and with the respective density of equilibrium trapped charge $Q_t(\mu) = eN_t(\mu)$ which is assumed to be positive for definiteness. For traps with small recharging time the equilibrium with the substrate would establish faster. These traps rapidly exchanged with the substrate are often referred as to the interface traps ($N_{it}$) [6], [7]. Defects which do not have time to exchange charge with the substrate during the measurement time (gate bias sweeping time) are referred to as oxide-trapped traps ($N_{ot}$). Difference between the interface and oxide traps is relative and depends, particularly, on the gate voltage sweep rate and the measurement's temperature. Interface trap capacitance per unit area $C_{it}$ may be defined in a following way

$$C_{it} \equiv \frac{d}{d\mu}(-eN_t(\mu)) > 0. \tag{10}$$

Note that the Fermi level dependent $eN_t(\mu)$ contains the charge on all traps, but for a finite voltage sweep time $t_s$ only the "interface traps" with low recharging time constants $\tau_t < t_s$ contribute to the recharging process.

Interface trap capacitance (F/cm$^2$) with accuracy up to the dimensional factor represents the energy density of the defect levels $D_{it}$ ( cm$^{-2}$eV$^{-1}$). It is easy to see that these values are related as

$$C_{it} = e^2 D_{it}(\mu). \qquad (11)$$

It is useful to note that 1 fF/μm$^2 \cong 6.25 \times 10^{11}$ cm$^{-2}$ eV$^{-1}$. The typical interface trap capacitance in modern silicon MOSFETs lies within the range $D_{it} \sim 10^{11}$ -$10^{12}$ cm$^{-2}$ eV$^{-1}$ and is rather sensitive (especially for thick (> 10 nm) insulated layers) to ionizing radiation impact [7].

## 4. General Electrostatics of GFET

Let us consider the simplest form of the gate-insulator-graphene (GIG) structure representing the two-plate capacitor capable to accumulate charges of the opposite signs. Without loss of generality we will reference the chemical potential in graphene from the level of charge neutrality $E_{NP}$. Electron affinity (or work function for Dirac point) of graphene with the reference of the vacuum energy level $E_{vac}$ can be defined as

$$\chi_g = E_{vac} - E_{NP}. \qquad (12)$$

Note that the graphene work function is of order of $\chi_g \sim 4.5$ eV [8]. It is well known that voltage bias between any device's nodes is equivalent to applying of electrochemical potential bias. There are generally at least two contributions to the electrochemical potential

$$\mu = \zeta + U = \zeta - e\varphi \qquad (13)$$

where $\zeta$ is proper electric charge independent chemical potential, $U$ and $\varphi$ are the electrostatic energy and potential $U= -e\varphi$. Neglecting voltage drop in the gate made routinely of good 3D conductors due to its extremely large quantum capacitance per unit area we get

$$\mu_{gate} = -e\varphi_{gate} - W_{gate}, \qquad (14)$$

$$\mu_{graphene} = -\chi_g + \zeta - e\varphi_{graphene} = E_{NP} + \zeta, \qquad (15)$$

where $\varphi_{graphene}$ is electrostatic potential of graphene sheet, $W_{gate}$ is work function of the gate material, and $E_{NP} = -\chi_g - e\varphi_{graphene}$ is the energy position of the charge neutrality (or, Dirac) point.

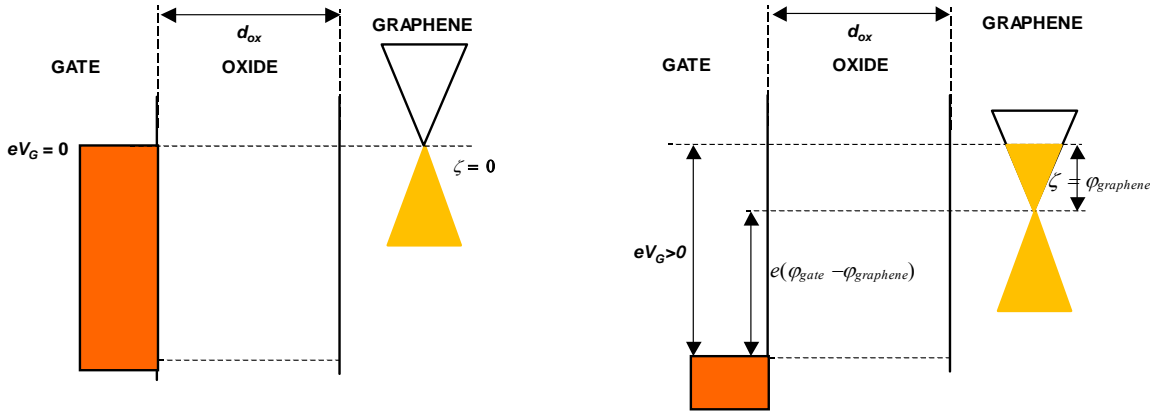

Fig. 2. Band diagram of gate–oxide- graphene structure at $V_G = 0$ (left) and $V_G > 0$ (right). Here, $\varphi_{gg} = 0$, for simplicity.

Applying the gate voltage (to say, positive) with reference of grounded graphene plate we increase the chemical potential and electrostatic potential of the graphene sheet so as they exactly compensate each other keeping the electrochemical potential of the graphene sample unchanged (see Fig. 2).

Particularly, the electrical bias between the metallic (or almost metallic) gate and the graphene sample is equal to a difference between the electrochemical potentials in graphene ($\mu_{graphene}$) and the gate ($\mu_{gate}$)

$$eV_G = \mu_{graphene} - \mu_{gate} = \varphi_{gg} + \zeta + e(\varphi_{gate} - \varphi_{graphene}). \qquad (16)$$

where $\varphi_{gg} \equiv W_{gate} - \chi_g$ is the work function difference between the gate and graphene. For zero oxide charge (or, for charged oxide defects located nearly the insulator-graphene interface) the electric field $E_{ox}$ is uniform across the gate thickness ($d_{ox}$) and one reads

$$\varphi_{gate} - \varphi_{graphene} = E_{ox} d_{ox} = \frac{eN_{gate}}{\varepsilon_{ox}\varepsilon_0} d_{ox} \equiv \frac{eN_{gate}}{C_{ox}}, \qquad (17)$$

where $N_{gate}(V_G)$ is the number of charge carriers on the metallic gate per unit area and the oxide (insulator) capacitance per unit area $C_{ox}$ expressed through the dielectric constants of the insulator ($\varepsilon_{ox}$) is defined as

$$C_{ox} = \frac{\varepsilon_{ox}\varepsilon_0}{d_{ox}}. \qquad (18)$$

## 5. Characteristic Scales of Gated Graphene

The planar electric charge neutrality condition for the total gated structure can be written down as follows

$$N_G + N_t = n_S, \qquad (19)$$

where $N_G$ is the number of positive charges per unit area on the gate; $n_s$ is the charge imbalance density per unit area ($n_s$ may be positive or negative and generally non-integer), $N_t$ is the defect density per unit area which is assumed to be positively charged (see Fig.3). Then total voltage drop (Eq.16) across the structure becomes modified as

$$eV_G = \varphi_{gg} + \varphi + \frac{e^2}{C_{ox}}\left(n_S(\zeta) - N_t(\zeta)\right). \qquad (20)$$

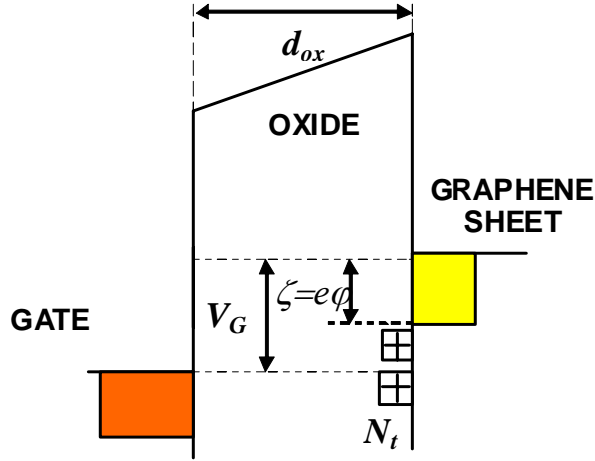

Fig. 3. Band diagram of graphene FET.

The voltage corresponding the electric charge neutrality point gate $V_{NP}$ is defined in a natural way

$$V_{NP} \equiv V_G(\zeta = 0) = \varphi_{gg} - \frac{eN_t(\zeta = 0)}{C_{ox}}. \qquad (21)$$

Chemical potential is positive (negative) at $V_G > V_{NP}$ ($V_G < V_{NP}$). Then we have

$$¶ e(V_G - V_{NP}) = \zeta + \frac{e^2 n_S}{C_{ox}} + \frac{e^2(N_t(\zeta = 0) - N_t(\zeta))}{C_{ox}}. \qquad (22)$$

Taking for brevity without loss of generality $V_{NP} = 0$ and assuming zero interface trap charge at the NP point as well as constant density of trap states we have

$$e^2\left(N_t(\zeta = 0) - N_t(\zeta)\right) \cong C_{it}\,\zeta. \qquad (23)$$

Taking into account Eq.23 the basic equation of graphene planar electrostatics can be written down in a form

$$eV_G = \varepsilon_F + \frac{e^2 n_S}{C_{ox}} + \frac{C_{it}}{C_{ox}}\varepsilon_F \equiv m\varepsilon_F + \frac{\varepsilon_F^2}{2\varepsilon_a}, \qquad (24)$$

where we have introduced for convenience a dimensionless "ideality factor"

$$m \equiv 1 + \frac{C_{it}}{C_{ox}}, \qquad (25)$$

and notation $\varepsilon_F$ used instead of $\zeta$. The specificity of the graphene-insulator-gate structure electrostatics is reflected in Eq.24 in appearance of the characteristic energy scale

$$\varepsilon_a = \frac{\pi\hbar^2 v_0^2 C_{ox}}{2e^2} = \frac{\varepsilon_{ox}}{8\alpha_G}\frac{\hbar v_0}{d_{ox}}, \qquad (26)$$

where the graphene "fine structure constant" is defined as ( in SI units)

$$\alpha_G = \frac{e^2}{4\pi\varepsilon_0 \hbar v_0}. \qquad (27)$$

This energy is nothing but the full electrostatic energy stored in the capacitor with the area fall at one carrier in graphene.

Fig.4 shows dependencies of characteristic electrostatic energy of gated graphene $\varepsilon_a$ vs gate oxide thickness for typical dielectric constants 4 (SiO$_2$) and 16 (HfO$_2$).

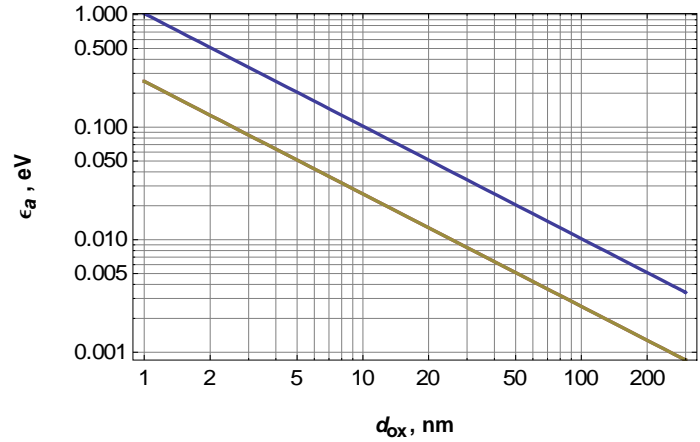

Fig. 4. The dependencies of the $\varepsilon_a$ as functions of the insulator thickness $d_{ox}$ for different dielectric permittivity equal to 4 (lower curve) and 16 (upper curve).

Energy scale $\varepsilon_a$ bring in a natural spatial scale specific to the graphene gated structures

$$a_Q \equiv \frac{\hbar v_0}{\varepsilon_a} = \frac{2e^2}{\pi\hbar v_0 C_{ox}} = \frac{8\alpha_G}{\varepsilon_{ox}}d_{ox}, \qquad (28)$$

and corresponding characteristic density

$$n_Q \equiv \frac{1}{\pi a_Q^2} = \frac{\varepsilon_a^2}{\pi\hbar^2 v_0^2} = n_S(\varepsilon_F = \varepsilon_a). \qquad (29)$$

Due to the fact that graphene "fine structure constant" $\alpha_G \cong 2.0 - 2.2$ the characteristic length $a_Q$ is occasionally of order of the oxide thickness for the insulators with $\varepsilon_{ox} \sim 16$ (i.e. for HfO$_2$). Interestingly that the energy scale $\varepsilon_a$ can be as well represented as functions of the Fermi energy and wavevector $k_F$, quantum capacitance and charge density

$$\kappa \equiv \frac{\varepsilon_a}{\varepsilon_F} = \frac{C_{ox}}{C_Q} = \frac{1}{k_F a_Q} = \frac{1}{\sqrt{\pi n_S a_Q^2}}, \qquad (30)$$

where $\kappa$ is defined independently as the ratio of the diffusion to the drift component in the channel [9].

## 6. Self-Consistent Solution of Basic Electrostatic Equation

Solving algebraic Eq. (24) one obtains an explicit dependence (to be specific for $V_G > 0$) of the electron Fermi energy as function of the gate voltage

$$\varepsilon_F = \left(m^2\varepsilon_a^2 + 2\varepsilon_a eV_G\right)^{1/2} - m\varepsilon_a \qquad (31)$$

This allows to immediately write the explicit relation for graphene charge density dependence on gate voltage

$$\frac{e^2 n_S}{C_{ox}} = eV_G - m\varepsilon_F = eV_G + m^2\varepsilon_a - m\left(m^2\varepsilon_a^2 + 2\varepsilon_a eV_G\right)^{1/2}. \qquad (32)$$

Restoring omitted terms the latter equation can be rewritten as [10], [11]

$$en_S(V_G) = C_{ox}\left(|V_G - V_{NP}| + V_0\left(1 - \left(1 + 2\frac{|V_G - V_{NP}|}{V_0}\right)^{1/2}\right)\right), \qquad (33a)$$

where the characteristic voltage $V_0 \equiv m^2\varepsilon_a / e$ is defined where interface trap capacitance is taken into account. One can modify Eq.33 taking into account the finite total carrier density $N_S(\mu = 0)$ at the charge neutrality point

$$en_S(V_G) = e\frac{\pi}{3}\left(\frac{k_B T}{\hbar v_0}\right)^2 + C_{ox}\left(|V_G - V_{NP}| + V_0\left(1 - \left(1 + 2\frac{|V_G - V_{NP}|}{V_0}\right)^{1/2}\right)\right). \qquad (33b)$$

The modified Eq.33b yields results almost identical to the formally exact Eq.6.

Figs. 5-6 exhibit numerically the interrelation of $V_0$ with $C_{it}$ and $d_{ox}$.

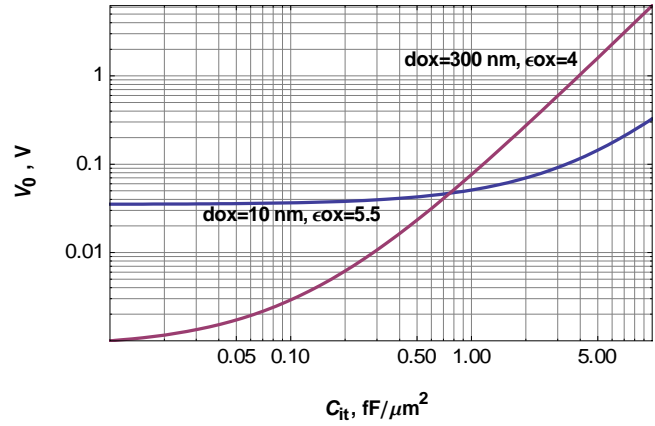

Fig. 5. Simulated dependencies of the characteristic voltage $V_0$ as functions of the interface trap capacitance $C_{it}$ for different oxide parameters.

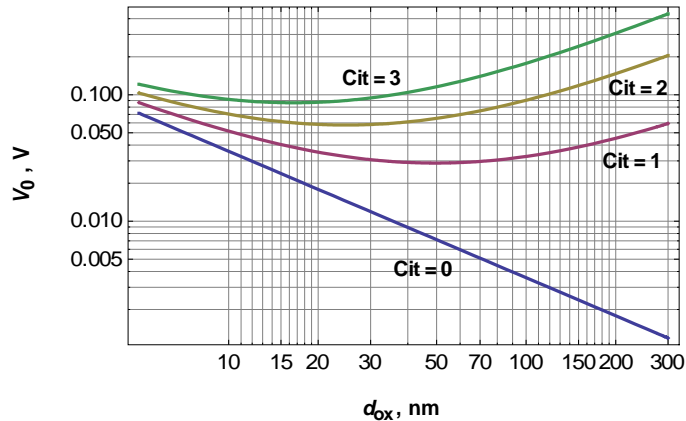

Fig. 6. Simulated dependencies of the characteristic voltage $V_0$ as functions of oxide thickness for different interface trap capacitance (in fF/μm²).

View of charge density dependence versus gate voltage is determined by relations of characteristic values (see Fig. 5, 6). At relatively high gate voltage $|V_G - V_{NP}| \gg V_0$ (or, the same, for "thick" oxide) we have close to linear dependence

$$en_S \cong C_{ox}\left(|V_G - V_{NP}| - \left(2V_0|V_G - V_{NP}|\right)^{1/2}\right). \tag{34}$$

Most part of external gate voltage drops in this case on the oxide thickness. Such is the case of "standard" oxide thickness $d_{ox}$ = 300 nm. Actually for not too small gate bias the charge density dependence on gate voltage is very close to linear [12]. For future graphene FET the gate oxide thickness is assumed to be of order of few or ten of nanometers. For such case of much thinner oxides or under relatively small gate biases $C_{ox}|V_G - V_{NP}| < en_Q$ we have quadratic law for density dependence (see Fig. 2b)

$$en_S \cong C_{ox}(V_G - V_{NP})\left(\frac{V_G - V_{NP}}{V_0}\right), \quad V_G - V_{NP} < V_0. \tag{35}$$

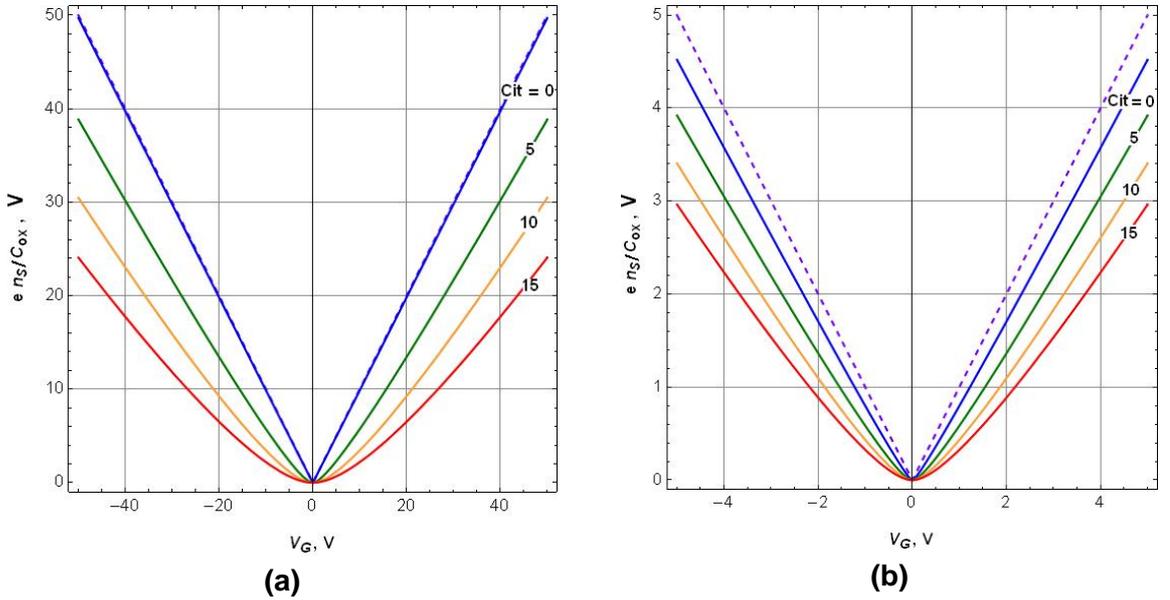

Fig. 7. Simulated charge density dependencies in reduced form $en_S/C_{ox}$ as functions of gate voltage for $\varepsilon_{ox}$ = 4 and different interface trap capacitance $C_{it}$ = 0, 5, 10, 15 $f$F/μm$^2$; (a) $d_{ox}$ = 300 nm; (b) $d_{ox}$ = 10 nm. Dashed curves correspond to $en_S/C_{ox} = V_G$.

Fig.7 show that $n_S(V_G)$ curves are strongly affected by interface trap recharging even for relatively thin oxides.

## 7. Gate and channel capacitance

Capacitance-voltage measurements are very important in providing information about gated field-effect structures. Taking derivative of Eq. 20 with respect to chemical potential, we have

$$\frac{dV_G}{d\mu} = 1 + \frac{C_Q + C_{it}}{C_{ox}}. \tag{36}$$

Low-frequency gate capacitance can be defined as

$$C_G = e\left(\frac{\partial N_G}{\partial V_G}\right) = e\frac{dN_G/d\mu}{dV_G/d\mu} = \frac{C_Q + C_{it}}{1 + \frac{C_Q + C_{it}}{C_{ox}}} = \left(\frac{1}{C_{ox}} + \frac{1}{C_Q + C_{it}}\right)^{-1} \tag{37}$$

Note that $C_G$ is often referred to as "total gate capacitance $C_{tot}$" in literature wherein the interface trap capacitance is frequently ignored. The Eq.37 corresponds to the equivalent electric circuit which is shown in Fig.8.

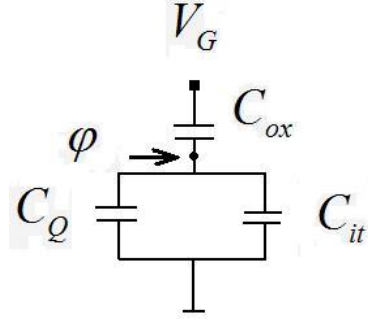

Fig. 8. Equivalent circuit of gated graphene.

One might introduce another relation corresponding to the intrinsic channel capacitance

$$C_{CH} = e\left(\frac{\partial N_S}{\partial V_G}\right) = e\frac{dN_S/d\mu}{dV_G/d\mu} = \frac{C_Q}{1+\frac{C_Q+C_{it}}{C_{ox}}} = \frac{C_{ox}}{1+\frac{C_{ox}+C_{it}}{C_Q}}, \quad (38)$$

where all capacitances are non-zero and assumed to be positive values for any gate voltage. The gate and the channel capacitances are connected in graphene gated structures through exact relation

$$\frac{C_G}{C_{CH}} = 1 + \frac{C_{it}}{C_Q} \quad (39)$$

and can be considered to be coincided only for ideal devices without interface traps when $C_{it} = 0$. All relationships for the differential capacitances remain valid for any form of interface trap energy spectrum. In an ideal case capacity-voltage characteristics $C_{CH}(V_G)$ should be symmetric with refer to the neutrality point implying approximately flat energy density spectrum of interface traps. For the latter case the channel capacity can be derived by direct differentiation of explicit dependence $n_S(V_G)$ in Eq.33

$$C_{CH} = e\frac{dn_S}{dV_G} = C_{ox}\left[1 - \frac{1}{\left[1 + 2|V_G - V_{NP}|/V_0\right]^{1/2}}\right]. \quad (40)$$

As can be seen in Fig.9 the capacitance-voltage characteristics $C_G(V_G)$ is strongly affected by the interface trap capacitance.

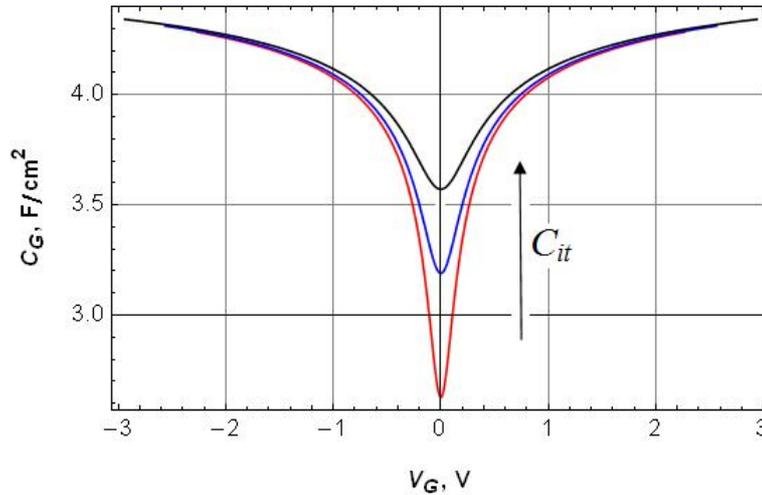

Fig. 9. Simulated dependencies of the gate capacitance $C_G(V_G)$ for different $C_{it} = 1, 5, 10$ $fF/\mu m^2$; $d_{ox} = 10$ nm, $\varepsilon_{ox} = 5.5$ (Al$_2$O$_3$).

For the case $C_{it} = 0$ (i.e. $m = 1$) capacitance-voltage dependencies can be considered as to be universal curves depending on only thickness and permittivity of the gate oxide through the parameter $\varepsilon_a$. In practice one should discriminate the quantum and the interface trap capacitances and this is a difficult task since they are in a parallel connection in equivalent circuit. Comparison of "ideal" capacitance –voltage characteristics with real measured ones represent a standard method of interface trap spectra parameter extraction [2], [13].

## 8. Experimental Data Analysis

Capacitance vs gate voltage dependencies in graphene gated structures have reported by several experimental groups followed by recalculation of quantum capacitance [14], [15], [16], [17], [18]. Although the measurements showed the expected V-shape dependence centered at the NP, the data, as a rule, are still far from ideal, making difficult to extract consistently the parameters of universal quantum capacitance dependence on the Fermi energy. Particularly, Fig.10 shows quantum capacitance dependencies adapted from independent results performed by the two independent experimental groups [14], [18].

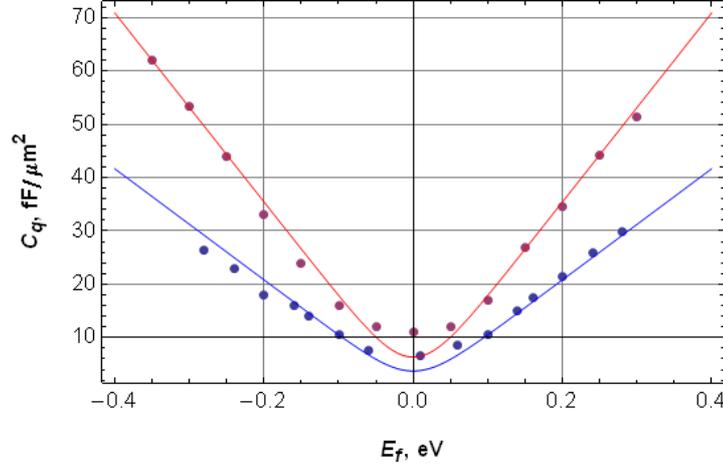

Fig. 10. Quantum capacitance vs the Fermi energy points recalculated from the capacitance data by the two experimental groups. The upper set of circles (red online) corresponds to the gate ($C_{ox}$ = 4.7 $fF/\mu m^2$ [18]), and lower circles (blue online) represent data of Ref. [14] ($C_{ox}$ = 5.6 $fF/\mu m^2$). These data are described by Eq. 9 with $v_0 \cong 1.15 \cdot 10^8$ cm/s for the upper curve (as obtained in original Ref. [18]) and $v_0 \cong 1.5 \cdot 10^8$ cm/s for the lower curve (as computed by us based on data in Ref. [14])

Despite of the both curves seem to be rather symmetric (especially, the upper) but they are obviously not coincident and admittedly far from a unique universal dependence which has to be described by ideal Eq.9. We argue here that the pointed disagreement follows from lack of consistent account of the interface trap capacitance under recalculation from initial measured capacitance data to quantum capacitance. Determination of the interface trap and quantum capacitances has to be considered as simultaneous and self-consistent procedure of their separation. Moreover the characteristic graphene velocity could be corrected in certain limits to adjust recalculated experimental quantum capacitance dependencies to the known universal curve.

## 9. Quantum and Interface Trap Capacitance Separation Procedure

If we have "ideal" structure with $C_{it} = 0$ (i.e. $m = 1$) then the gate capacitance depends only on a single dimensional parameter $\varepsilon_a$ containing the gate oxide capacitance and the characteristic graphene velocity $v_0$. In practice one should discriminate quantum interface trap capacitance and this is a difficult task since it is in parallel connection with the interface trap capacitance in an equivalent electric circuit.

A following iteration procedure can be used for separation. At first we set $C_{it} = 0$ ($m = 1$) and $C_G = C_{CH}(1 + C_{it}/C_Q) \simeq C_{CH}$ (recall that $C_{Qmin} \leq 10\, fF/\mu m^2$ at room temperatures). Then one can replot the experimental data for $C_G(V_G)$ using a reformulation of Eq.40

$$|V_G - V_{NP}| = \frac{V_0}{2}\left(\frac{1}{(1 - C_{CH}/C_{ox})^2} - 1\right) \cong \frac{V_0}{2}\left(\frac{1}{(1 - C_G/C_{ox})^2} - 1\right). \qquad (41)$$

Fig.11 shows a typical result of such graphical representation of experimental data which turns out to be linear in full agreement with Eq.41.

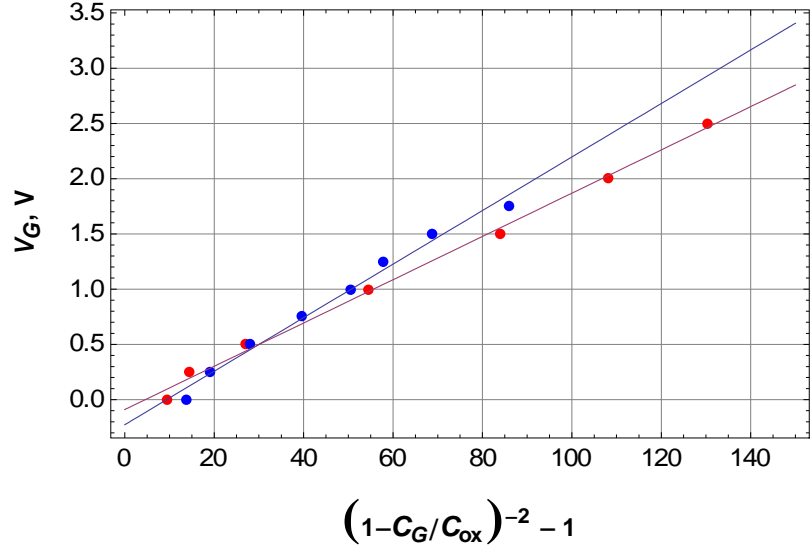

Fig. 11. Replot of the capacitance data illustrating the finding of $C_{it}$ with least-squares method. Upper line (blue circles online) corresponds to data in Ref. [13], the lower (red circles online) line corresponds to data in Ref. [17].

The slope of this linear dependence yields an experimental value of the characteristic voltage $V_0$ which depends on $C_{it}$. If we were aware exactly $v_0$ and $C_{ox}$ one could immediately to obtain $m$ and $C_{it}$. The ratio of the slopes does not depend on $v_0$ (indices 1 (2) correspond to the data in Ref. [14] (Ref. [18]))

$$\frac{V_0(1)}{V_0(2)} = \frac{m_1^2}{m_2^2}\frac{C_{ox1}}{C_{ox2}} \qquad (42)$$

and can be determined immediately from the Fig.11 $V_0(1)/V_0(2) = 1.237$. Using the known oxide capacitances we have found the ratio $m_1/m_2 = 1.02$ and $C_{it1}/C_{it2} = 1.53$.

Setting as a zero approximation $v_0 = 1.15\ 10^8$ cm/s (found in Ref.[18]) one can compute $C_{it}$ which turn to be non-equal but both of order 1 fF/μm$^2$ for results of the both experimental groups [14, 18] in contrast to own values used by the authors. Recall, that the values $C_{it} = 10$ $f$F/μm$^2$ [14] and $C_{it} = 0$ [18] were used at data treatment as far as we know. Furthermore with a use Eqs.37 the first iteration for quantum capacitance as function of Fermi energy can be calculated through experimental data

$$C_Q = \left(\frac{1}{C_G} - \frac{1}{C_{ox}}\right)^{-1} - C_{it}. \qquad (43)$$

Recalculated in this manner experimental points for $C_Q$ are found to be lower than original results in [18] where interface trap capacitance were ignored and to be higher than in [14] where $C_{it}$ were overestimated. In addition the experimental points from independent data have laid practically on a single curve, corresponding to characteristic graphene velocity in the range $1.15 \times 10^8$ cm/s $< v_0 < 1.5 \times 10^8$ cm/s.

Iterating the procedure for self-consistency we have found a following set of best fit parameters represented in Table 1.

*Table 1*

| Reference | $C_{ox}$, $f$F/μm$^2$ | $m$ | $C_{it}$, $f$F/μm$^2$ | $v_0$, $10^8$ cm/s |
|---|---|---|---|---|
| [Chen et al., 2009] | 5.6 [14] | 1.10 | 0.55 | 1.30±0.05 |
| [Ponomarenko et al., 2010] | 4.7 [18] | 1.08 | 0.36 | 1.30±0.05 |

Comparison of original and recalculated by us dependencies shown in Fig.12 exhibits the fact that consistent account of the interface trap capacitance is a necessary condition for obtaining of universal parameter of quantum capacitance.

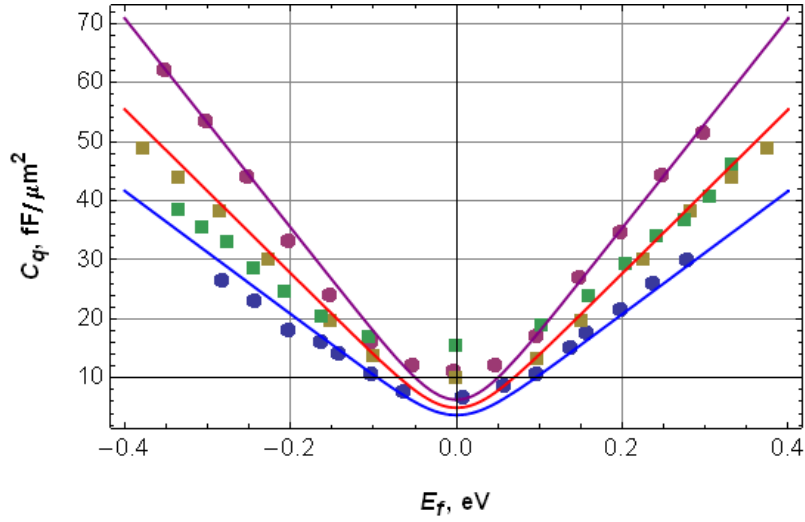

Fig. 12. Quantum capacitance curves recalculated as functions of the Fermi energy. Upper (purple) curve corresponds to [18] data and simulation with $v_0 = 1.15 \times 10^8$ cm/s (with ideal $C_Q$ dependence). The lower (blue circles online) curve corresponds to [14] which is simulated by us with $v_0 = 1.5 \times 10^8$ cm/s (in fact from Fig.8. of [14]). Taking into account extracted interface capacitances the experimental points of both groups have laid on a single medium (red) curve with $v_0 = 1.3 \times 10^8$ cm/s (yellow squares, [18]; green squares [14]).

Notice that found interface trap capacitances correspond numerically to a reasonable range of interface trap density of states $D_{it} \cong (2.2 - 3.4) \times 10^{11}$ cm$^{-2}$ eV$^{-1}$ typical for pristine gate oxides in the silicon MOSFETs.

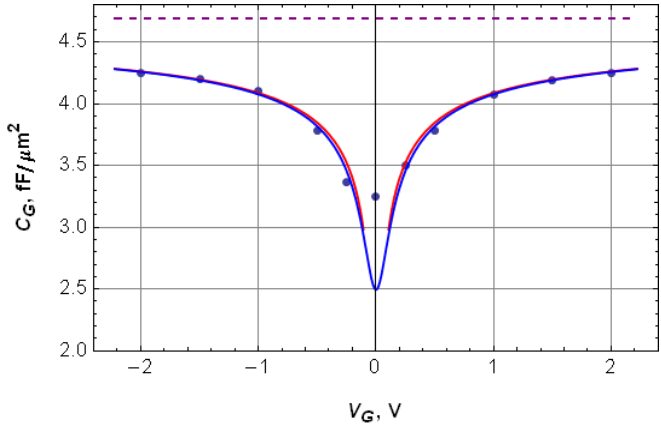 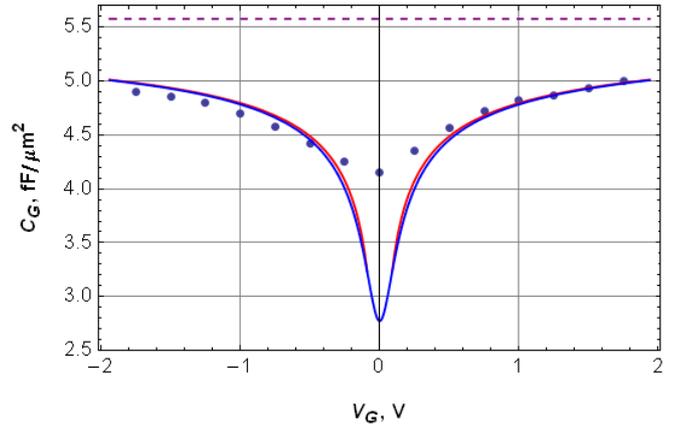

Fig. 13a. Comparison of the experimental gate capacitance dependence $C_G(V_G)$ obtained in [18] (points) and simulation with the Eqs.37 (blue) and 40 (red) (taking into account Eq.39). Used constants are $v_0 = 1.3 \times 10^8$ cm/s, $C_{ox} = 4.7$ $f$F/μm$^2$ [18], $C_{it}$ (extracted) = 0.36 $f$F/μm$^2$. The dashed curve shows $C_{ox}$.

Fig. 13b. Comparison of the experimental gate capacitance dependence $C_G(V_G)$ obtained in [14] (points) and simulation with the Eqs.37 (blue) and 40 (red) (taking into account Eq.39). Used constants are $v_0 = 1.3 \times 10^8$ cm/s, $C_{ox} = 5.6$ $f$F/μm$^2$ [14], $C_{it}$ (extracted) = 0.55 $f$F/μm$^2$ (unlike to $C_{it}$ = 10 $f$F/μm$^2$ used in [14]).

Fig.13 (a and b) shows the experimental gate capacitance as functions of gate voltage obtained in papers [18] and [14]. Based on formulas for homogeneous graphene the gate capacitance estimations strongly underestimate the capacitance values nearby the charge neutrality point. As can be seen in Figs.13 the differences between the experimental and calculated values correlate with the interface trap density: the greater disorder and concerned with it the interface trap capacitance the greater underestimation for homogeneous graphene approximation. It is known that due to occurrence of potential fluctuation induced by the charged oxide defects, graphene charge sheet breaks near the neutrality point into electron and hole "puddles" [19, 20]. This electron-hole puddles are capable to significantly increase the minimal quantum capacitance value that may yields independent information about charged defects density trapped in the insulator near the graphene sheet.

## 10. Electron-hole puddles and quantum capacitance

The potential fluctuation induced by charged near-interfacial defects distributed in uncorrelated way in the insulator can be described by Gaussian distribution function

$$P(u) = \frac{1}{\sqrt{2\pi \langle \delta u^2 \rangle}} \exp\left(-\frac{u^2}{2\langle \delta u^2 \rangle}\right) \tag{44}$$

where $u$ is fluctuating potential reckoning from a mean value, $\langle \delta u^2 \rangle$ is the dispersion of potential fluctuation. The standard deviation for potential of uncorrelated near-interfacial defects can be assessed as [21]

$$\langle \delta u^2 \rangle = \frac{e^4}{(4\pi\varepsilon_0 \bar{\varepsilon})^2} \pi n_{imp} \tag{45}$$

and to be determined by a sum of the positively and negatively charged defect densities $n_{imp} = n_{imp}^{(+)} + n_{imp}^{(-)}$; $\bar{\varepsilon}$ is a half-sum of the permittivities for the dielectrics adjusted to the graphene sheet. In the Thomas-Fermi approximation the local value of charge density is

$$n_S = sgn(\varepsilon_F - u(\mathbf{r})) \frac{(\varepsilon_F - u(\mathbf{r}))^2}{\pi \hbar^2 v_0^2} \tag{46}$$

where $\varepsilon_F$ is a single equilibrium Fermi energy of the inhomogeneous system.

At first we have to calculate the total net electric charge in graphene as function of $\varepsilon_F$ taking into account occurrence of potential fluctuation and electron-hole puddles

$$Q^{(p)}(\varepsilon_F) \cong \frac{e}{\pi \hbar^2 v_0^2} \left( \int_{-\infty}^{\varepsilon_F} (\varepsilon_F - u)^2 P(u) du - \int_{\varepsilon_F}^{\infty} (u - \varepsilon_F)^2 P(u) du \right) =$$

$$= \frac{e}{\pi \hbar^2 v_0^2} \left[ (\langle \delta u^2 \rangle + \varepsilon_F^2) erf\left(\frac{\varepsilon_F}{\sqrt{2\langle \delta u^2 \rangle}}\right) + \varepsilon_F \sqrt{\langle \delta u^2 \rangle} \sqrt{\frac{2}{\pi}} \exp\left(-\frac{\varepsilon_F^2}{2\langle \delta u^2 \rangle}\right) \right]. \tag{47}$$

Then the quantum capacitance accounting the electron-hole puddles becomes

$$C_Q^{(p)} = e\frac{\partial Q}{\partial \varepsilon_F} = \frac{2e^2 \varepsilon_F}{\pi \hbar^2 v_0^2} \left[ erf\left(\frac{\varepsilon_F}{\sqrt{2\langle \delta u^2 \rangle}}\right) + \frac{\sqrt{\langle \delta u^2 \rangle}}{\varepsilon_F} \sqrt{\frac{2}{\pi}} \exp\left(-\frac{\varepsilon_F^2}{2\langle \delta u^2 \rangle}\right) \right]. \tag{48}$$

The latter relation can be obtained immediately by direct averaging of density of states $C_Q^{(p)} = (e^2/\pi \hbar^2 v_0^2) \int_{-\infty}^{\infty} |\varepsilon - u| P(u) du$. The Eqs. 47 and 48 do not contain temperature since to be only valid for a conditions $k_B T < |\varepsilon_F| < \langle \delta u^2 \rangle^{1/2}$ or $k_B T < \langle \delta u^2 \rangle^{1/2} < |\varepsilon_F|$.

At the CNP we have the minimum quantum capacitance in disordered graphene

$$C_{Q\min}^{(p)} = C_Q^{(p)}(\varepsilon_F = 0) = \frac{2e^2}{\pi \hbar^2 v_0^2} \sqrt{\frac{2\langle \delta u^2 \rangle}{\pi}}, \tag{49}$$

which determines an observed plateau in quantum capacitance dependencies for $|\varepsilon_F| < \langle \delta u^2 \rangle^{1/2}$. Figure 14 shows the comparison of the experimental and simulated gate capacitance characteristics obtained with corrected quantum capacitance Eq. 48 by fitting of potential standard deviation.

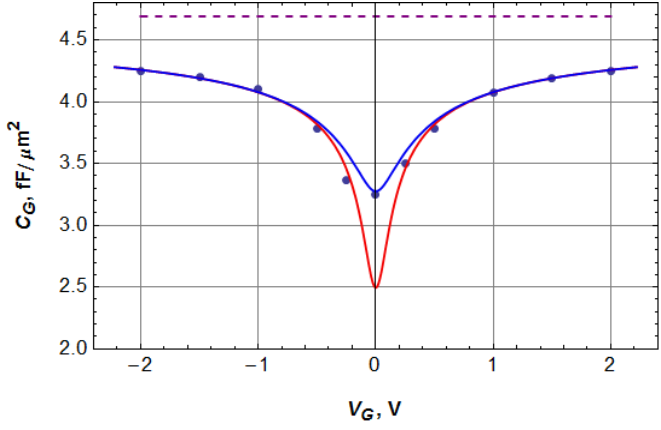 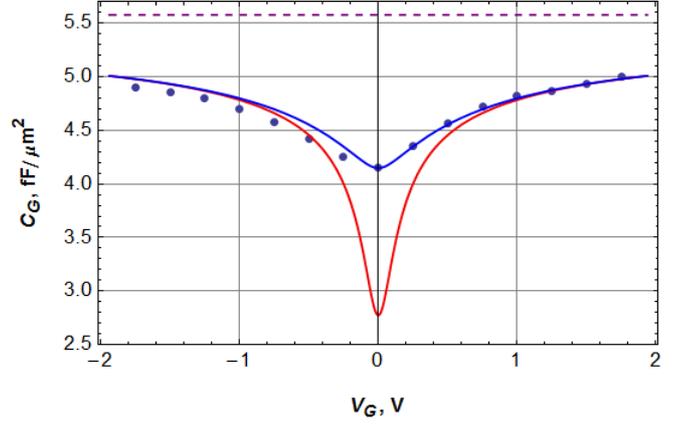

Fig.14a. Simulated gate capacitance dependence in comparison with experimental points [17]. Fitted value of potential dispersion is $\langle \delta u^2 \rangle^{1/2} = 95$ meV. The lower curve corresponds to $\langle \delta u^2 \rangle^{1/2} = 0$. All other parameters are taken the same as in Fig.13.

Fig.14b. Simulated gate capacitance dependence in comparison with experimental points [13]. Fitted value of potential dispersion is $\langle \delta u^2 \rangle^{1/2} = 141$ meV. All other parameters are taken the same as in Fig.13.

The charged impurity total concentrations $n_{imp}$ computed with Eq.45 are $3.9 \times 10^{12}$ cm$^{-2}$ and $8.6 \times 10^{12}$ cm$^{-2}$ for fitted standard deviations 95 and 141 meV. Recalculated quantum capacitance curves corrected with account of electron-hole puddle contribution are depicted in Fig.15. Eq.48 was used in Fig.15 instead of Eq.9 for ideal graphene.

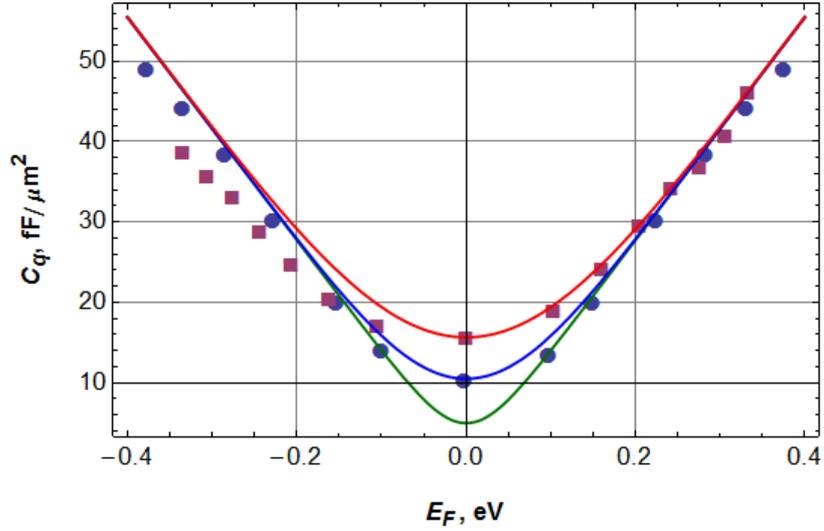

Fig.15. Corrected quantum capacitance curves recalculated for the two experiments taking into account electron-hole puddles. Red (blue) curve corresponds to data taken from [Chen, 2008]([Ponomarenko, 2010]). Lower (green) curve corresponds to the universal relation (Eq.9) for ideal graphene. All other parameters are taken the same as in Figs. 13-14.

## 11. Conductivity averaging

The conductivity in graphene is given by the conventional formula $\sigma_0 = e \mu_0 N_S$, where $\mu_0$ is the mobility and $N_S$ is the total concentration of carriers of both signs. First, we have calculated the total carrier (electron + hole) density in inhomogeneous graphene sheet

$$N_S^{(p)}(\varepsilon_F) \cong \frac{e}{\pi \hbar^2 v_0^2} \left( \int_{-\infty}^{\infty} (\varepsilon_F - u)^2 P(u) du \right) = \frac{\langle \delta u^2 \rangle + \varepsilon_F^2}{\pi \hbar^2 v_0^2} . \tag{50}$$

Obviously, the Eq.50 implies that the residual carrier concentration at the CNP [22] is determined immediately by the potential fluctuation dispersion

$$n_0 = N_S^{(p)}(\varepsilon_F = 0) = \frac{\langle \delta u^2 \rangle}{\pi \hbar^2 v_0^2} = \frac{\alpha_G^2}{\bar{\varepsilon}_{ox}^2} n_{imp} \tag{51}$$

This is exactly the result, which can be derived with the Shklovskii argument of nonlinear screening [23] with the optimal size of the puddles and residual concentration

$$R_0 \cong \frac{\bar{\varepsilon}_{ox}^2}{\alpha_G^2 \sqrt{\pi n_{imp}}}, \qquad n_0 \cong \frac{\sqrt{\pi n_{imp} R_0^2}}{\pi R_0^2}. \tag{52}$$

Similar arguments has been used in Refs. [24] for description of disorder at the Si-SiO$_2$ interface in the silicon MOSFETs.

To assess the carrier mobility $\mu_0 = e v_0 \tau(\varepsilon_F)/p_F$ the relaxation time can be estimated through the Fermi Golden Rule

$$\frac{1}{\tau_{tr}(\varepsilon_F)} \sim \frac{2\pi}{\hbar} n_{imp} |u_{imp}|^2 g_{2D}(\varepsilon_F) = \frac{2\pi}{\hbar} n_{imp} |u_{imp}|^2 \frac{C_Q}{e^2}, \tag{53}$$

where $|u_{imp}|$ is the average of the scattering potential matrix element in 2D momentum space. According to Ref. [9] the screened matrix element is expressed as

$$|u_{imp}| = \frac{e^2}{C_Q + C_{ox}}, \tag{54}$$

where the gate screening of the Coulomb scatterers dominates near the CNP at $C_Q \ll C_{ox} = \varepsilon_{ox} \varepsilon_0 / d_{ox}$. Recall that $C_{ox}/C_Q = \varepsilon_a / \varepsilon_F$. Then in homogeneous case we find for conductivity

$$\sigma_0 = C \frac{e^2}{h} \frac{n_S}{n_{imp}} \left(1 + \frac{\varepsilon_a}{\varepsilon_F}\right)^2 = C \frac{e^2}{h n_{imp}} \frac{(\varepsilon_a + \varepsilon_F)^2}{\pi \hbar^2 v_0^2}. \tag{55}$$

Numerical factor $C$ ($\cong$ 1-30) should be calculated depending on generally unknown positions of the oxide-trapped charged defects.

Local conductivity in graphene with the long-range inhomogeneities can be expressed then through the local value of the chemical potential

$$\sigma[u(\mathbf{r})] = C \frac{e^2}{h} \frac{n_S[\varepsilon_F - u(\mathbf{r})]}{n_{imp}} \left(1 + \frac{\varepsilon_a}{\varepsilon_F - u(\mathbf{r})}\right)^2 = C \frac{e^2}{h n_{imp}} \frac{(\varepsilon_a + \varepsilon_F - u(\mathbf{r}))^2}{\pi \hbar^2 v_0^2}. \tag{56}$$

Notice, that local conductivity defined by Eq.56 remains finite even along "the p-n junction lines" of the Dirac points defined by the condition $\varepsilon_F = u(\mathbf{r})$.

Performing averaging $\sigma_0^{(p)} = \int_{-\infty}^{\infty} \sigma(u) P(u) du$ we obtain the mean low-field conductivity in the inhomogeneous graphene with the electron-hole puddles

$$\sigma_0^{(p)} = C \frac{e^2}{h n_{imp}} \frac{(\varepsilon_a + \varepsilon_F)^2 + \langle \delta u^2 \rangle}{\pi \hbar^2 v_0^2} \tag{57}$$

The minimum conductivity occurs at $\varepsilon_F = 0$

$$\sigma_{0\min}^{(p)} = C \frac{e^2}{h n_{imp}} \frac{\varepsilon_a^2 + \langle \delta u^2 \rangle}{\pi \hbar^2 v_0^2} = C \frac{e^2}{h} \frac{n_Q + n_0}{n_{imp}}. \tag{58}$$

For the illustration of our approach we have examined the experimental results presented in Ref. [25] where the resistivity $\rho = 1/e\mu_0 N_S$ was measured as function of gate voltage for different graphene samples. Due to phonon contribution into scattering we have simulated these results using experimental mobilities extracted by the authors of this paper. The dependences of total carrier concentration were computed as function of Fermi energy according Eq.6 $N_S = n_e + n_h + n_0$ with added constant residual concentration as a fitted parameter. Taking into consideration Eq.57 the substitution $\varepsilon_F \to \varepsilon_F + \varepsilon_a$ were in fact used in Eq.6 that practically had not influenced on the simulation results in this case due to smallness of $\varepsilon_a$ in thick gate oxides. A use of exact relation Eq.6 instead of approximate Eq.7 is essentially for smooth description of resistivity near the resistivity maximum at low temperatures.

The dependences of the Fermi energy $\varepsilon_F$ as functions of gate voltage were modeled with Eq.31. Such computation scheme allows to easily describe the impact of interface trap recharge on the shape of measured characteristics as functions of external gate voltage. Excepting trivial $V_{NP}$, we fit only the interface trap capacitance $C_{it}$ and the constant residual concentration $n_0$ for three different samples. The total charged defect density $n_{imp}$ then has been recalculated using Eq.51.

Comparison of experimental and simulated results is shown in Fig.16.

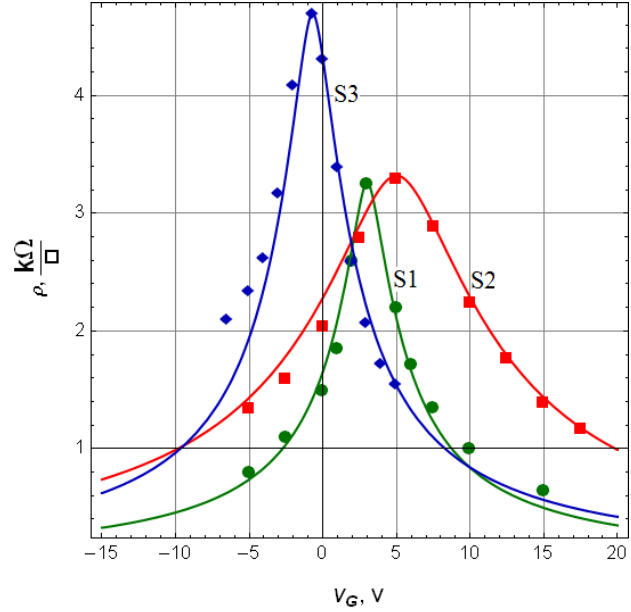

Fig. 16. The gate voltage dependence of the resistivity for the samples S1, S2, S3 [24]: the carrier mobilities were taken from [24] $\mu$ (S1) = 17500 cm$^2$/Vs, $\mu$ (S2) = 9300 cm$^2$/Vs, and $\mu$ (S3) = 12500 cm$^2$/Vs. Fitting results: the sample S1: $n_0$ = 0.67×10$^{11}$ cm$^{-2}$, $C_{it}$ = 3.5 fF/μm$^2$; the sample S2: $n_0$ = 1.6×10$^{11}$ см$^{-2}$, $C_{it}$ = 9.3 fF/μm$^2$; the sample S3: $n_0$ = 0.64×10$^{11}$ см$^{-2}$, $C_{it}$ = 4.4 fF/μm$^2$ (T = 200K, $\varepsilon_{ox}$ = 4, $d_{ox}$ = 300 nm, $v_0$ = 1.3×10$^8$ cm/s).

The extracted parameters for different samples are summarized in Table 2.

*Table* 2

| sample | $C_{it}$, fF/μm$^2$ | $n_0$, 10$^{11}$ cm$^{-2}$ | $n_{imp}$, 10$^{11}$ cm$^{-2}$ | $\mu_0$, cm$^2$/Vs [25] |
|---|---|---|---|---|
| S1 | 3.6 | 0.67 | 1.5 | 17500 |
| S2 | 9.5 | 1.6 | 3.5 | 9300 |
| S3 | 4.4 | 0.63 | 1.4 | 12500 |

Simulation results exhibit an excellent agreement with the experiment in description of vicinity of the "Dirac peak" for resistivity at reasonable values of extrinsic physical parameters. This suggests that the widths of Dirac peaks are determined mainly by the interface trap density. Behavior of the resistivity dependences at large |$V_G$ − $V_{NP}$| (where $\rho \leq 1$ kΩ/□) is typically influenced by the contacts.